\input amssym.def
\input amssym.tex

\documentstyle[12pt,a4,cite]{article}
\begin{document}
\pagestyle{headings}
\def\o3{$O(3)$}%
\def\eps{\epsilon}%
\def\su2{$SU(2)$}%
\def\gb{\overline{g}}%

\def\thefootnote{\fnsymbol{footnote}}%
\addtocounter{footnote}{1}
\def\appendix{\par
  \setcounter{section}{0}
  \setcounter{subsection}{0}
  \def\thesection{Appendix \Alph{section}}}

\begin{flushright}
Liverpool Preprint: LTH 331\\
hep-lat/yymmppp\\
April 5th 1994
\end{flushright}
\vspace{5mm}

\begin{center} 

{\LARGE\bf Instanton size distribution in \o3}\\[1.5cm]

{\bf C. Michael and  P.S. Spencer}\\
DAMTP, University of Liverpool, Liverpool L69~3BX, UK\\
\end{center}

\begin{abstract}
We present calculations of the size distribution of instantons in the 
$2d$ \o3 non-linear $\sigma$ model, and briefly discuss the effects cooling
has upon the configurations and the topological objects.
\end{abstract}

\section{Introduction}
The \o3 non-linear $\sigma $-model in $(1+1)$ space-time dimensions
has been much studied because of the properties it shares with
4-dimensional non-Abelian gauge theories: it is asymptotically free,
becomes non-perturbative in the infra-red regime and has a non-trivial
topology due to the homotopy class
\mbox{$\pi_{2}(S_{2})=\Bbb{Z}$} of windings from $S_{2}\rightarrow S_{2}$.

It has been known for some time that the standard lattice 
discretisation of the \o3 model has problems---for example, the lattice
topological susceptibility does not obey na\"{\i}ve scaling laws.
L\"{u}scher has shown~\cite{Lu:scale} that this is due to the model
being dominated by short-range fluctuations. Formulating this
theory on the lattice implies a minimum size to these fluctuations, and
so the important contributions from fluctuations smaller than one lattice
spacing are absent. There is also the problem that additional
short-range fluctuations (known as lattice artefacts) are present in the
lattice formulation and not in the continuum. These are unphysical field
configurations such as the winding of the field around one plaquette,
giving rise to spurious contributions to the topological
charge~\cite{Lu:scale}.

The model is well understood in some respects---for instance an exact
expression for the mass gap is known~\cite{massgap}.  It is thus a
natural candidate for a theory which one can attempt to  understand in
terms of the vacuum structure.  In this paper we explore the nature of
the vacuum via numerical simulation in Euclidean time. We concentrate 
on large scale structures in the vacuum - such as those arising from 
instantons. In order to focus on such extended features in the vacuum,
we use a procedure to smooth out local fluctuations.  A cooling
algorithm is used to achieve this.  The basic idea is that a local
smoothing of the fields should preserve long range features such as
instantons.  In~\cite{US:cool} we explored this by varying the details
of the cooling procedure and presenting qualitative features of the 
resulting topological charge distribution.

In this work we wish to make a quantitative study of the nature of the 
topological distribution in the vacuum. Such a cooling procedure will 
modify the short distance properties of the vacuum, so we explore  the
distribution of the size of topological objects since we can  expect
that the large size objects are unchanged under cooling. By varying the
lattice spacing and the extent  of cooling, we are able to present
results for the physical distribution  of large scale topological
fluctuations.

\smallskip
\section{The \o3 Model on a Lattice}\label{s:o3}
\par
The continuum $2d$ \o3 Euclidean action $S_{E}$ is defined as: 
\begin{equation}\label{e:contact}
S_{E}=\frac{1}{2g^{2}}\int d^{2}x(\partial_{\mu}\phi)^{2} 
\end{equation}
\noindent with $\phi$ an \o3 vector and the constraint
$\phi^{2}=1$.

Much use has been made of so-called ``improved'' actions in
\o3 (eg~\cite{Pisa:o3,Arisue:massgap}) and more recently
``over-improved''~\cite{Sp:overimp} and ``perfect''\cite{Has:perfect}
lattice actions have been proposed; here we use the more local
nearest-neighbour action. This is appropriate since we wish to develop
techniques that will extend to other theories (such as 4-dimensional
$SU(N)$) where a perfect lattice action is not yet available. Taking
the discrete form of the derivative ($\mu$ running over the space-time
directions):

\begin {equation}\label{e:findiff}
\partial_{\mu}\phi(x)\rightarrow \Delta_{\mu}\phi_{x}=
\frac{1}{a}(\phi_{x+\mu}-\phi_{x})
\end{equation}
\noindent where $\phi_{x}$ is now a field associated with the site $x$
and $a$ is the lattice spacing (hereafter set to 1), the lattice
action is:
\begin{eqnarray}\label{e:lattact}
S_{L}&=&\frac{1}{2g^{2}}\sum_{x, \mu}(\phi_{x+\mu}-\phi_{x})^{2}\\
     &=&\frac{1}{g^{2}}\sum_{x}(2-\phi_{F}(x)\cdot\phi_{x})\nonumber
\end{eqnarray}
\noindent with the index $\mu$ again running over directions on the
lattice, and $\phi_{F}$ defined as the sum over nearest neighbours:
\begin{equation}\label{e:phiF}
\phi_{F}(x)=\sum_{\mu}\phi_{x\pm\mu}.
\end{equation}

The continuum topological charge $Q$ is defined as:
\begin{equation}\label{e:ftq}
Q^T_f=\frac{1}{8\pi}\int d^{2}x \eps_{\mu\nu}\eps_{ijk}
\phi_{i}\partial_{\mu}\phi_{j}\partial_{\nu}\phi_{k}
\end{equation}
\noindent and is by definition an integer. However, if this ``field
theoretical'' definition is placed na\"{\i}vely onto the lattice, it
returns non-integer values and acquires a renormalisation
factor~\cite{Pisa:o3,Pisa:renorm}. There are other ``geometric''
lattice definitions that return integers, one example being given
in~\cite{Lu:scale,BL:o3stats,Berg:o3}, based on the mapping
$S_{2}\rightarrow S_{2}$, by calculating the signed area of the
spherical triangles formed by the $\phi$ fields around a
plaquette. The contribution to the total charge from a site $x^{*}$ on the
dual lattice is given by:
\begin{equation}
Q^T_g(x^{*})=\frac{1}{4\pi}\bigl( (\sigma
A)(\phi_{1},\phi_{2},\phi_{3})+(\sigma
A)(\phi_{3},\phi_{4},\phi_{1})\bigr)
\end{equation}
\noindent where 
\begin{equation}
\sigma(\phi_{1},\phi_{2},\phi_{3})={\rm
sign}(\phi_{1}\cdot\phi_{2}\times\phi_{3})
\end{equation}
\noindent and $A(\phi_{1},\phi_{2},\phi_{3})$
 is the area of the spherical triangle on the unit sphere mapped out by
the $\phi$ fields, and the subscripts $1,2,3,4$ refer to the corners of
a plaquette labelled anti-clockwise. Clearly there are two ways to 
triangulate a plaquette and the difference in the contribution to $Q^T$
from the two triangulations is $0$ for configurations with small action,
but can be $\pm 1$ for other field configurations.
 
There are arguments against the use of such a charge definition,
particularly in calculations of the topological susceptibility,
$\chi_t$, in $CP^{N-1}$ theories with small values of $N$. Campostrini
et al have shown~\cite{Pisa:smallN} that for values of $N=10$ and
larger, the geometrical formulation gives a sensible definition of
$\chi_t$, but not for lower values of $N$. 

\section{Instantons in \o3}\label{s:inst}
Since we shall wish to compare fluctuations in the topological charge 
density with instantons, we now discuss the details of the instanton 
contributions. Let us first summarise the results for the  continuum in an
infinite space-time region. 

With $\phi$ written in terms of the projective fields $\omega$,
$\overline{\omega}$: 
\begin{equation}\label{e:phi}
\phi_{1}=\frac{\omega+\overline{\omega}}{\omega\overline{\omega}+1}\;,\;
\phi_{2}=\frac{\omega-\overline{\omega}}{{\rm i}(\omega\overline{\omega}+1)}\;,\;
\phi_{3}=\frac{\omega\overline{\omega}-1}{\omega\overline{\omega}+1}
\end{equation}
\noindent 
(1--3 here are \o3 indices) a field configuration {\em representing} a
single (continuum) instanton of size $|\rho|$ at position $r$ can
be explicitly constructed
via~\cite{BL:o3stats,Berg:o3,MW:sph,P:strings,BukLip:I-AI}
\begin{equation}\label{e:oneinst}
\omega=\frac{\rho}{z-r}
\end{equation}
\noindent with $z=x+it$ the coordinate in the complex plane. This can
be easily extended to generate multiple instanton configurations; an
$N$-instanton configuration has the form:
\begin{equation}\label{e:multinst}
\omega=\sum_{i=1}^{N}\frac{\rho_{i}}{z-r_{i}}
\end{equation}
and has topological charge $N$. Similarly, a multi--anti-instanton
configuration with $Q^T=-N$ can be generated
via~\cite{P:strings,BukLip:I-AI}:
\begin{equation}\label{e:multanti}
\omega=\sum_{i=1}^{N}\frac{\overline{\rho}_{i}}{\overline{z}-\overline{r}_{i}}
\end{equation}
\noindent It should be noted that a configuration containing both
instantons and anti-instantons is not strictly a solution of the
classical equations of motion. In~\cite{BukLip:I-AI}, Bukhvostov and
Lipatov construct a general configuration containing $N^I$ instantons
and $N^A$ anti-instantons via
\begin{equation}\label{e:inst-anti}
\omega=\left(\sum_{i=1}^{N^I}\frac{\rho^I_i}{z-r^I_i}\right)
\left(\sum_{j=1}^{N^A}\frac{\overline{\rho}^A_j}{\overline{z}-\overline{r}^A_j}\right)
\end{equation}
\noindent based on the assumption that the \mbox{(anti-)}instantons have
sizes small compared to their separations, $\rho^I_i,\rho^A_j\ll\ \mid
r^I_i - r^A_j \mid$, or essentially different scales, e.g. $\rho^I_i
\ll \rho^A_j$, i.e. on the assumption that the instantons and
anti-instantons are weakly interacting. In this scenario, the \o3
model becomes equivalent to an exactly soluble fermion model.

With $\phi$ and $\omega$ as in eqs.~\ref{e:phi} and~\ref{e:oneinst}
above, the action and topological charge density become:
\begin{equation}\label{e:sandq}
 S(z)\:,\:Q(z)\propto\frac{\rho^{2}}{(\rho^{2}+\mid z-r \mid^{2})^{2}}.
 \end{equation}
\noindent where here the size parameter $\rho$ is an abbreviation  for
$|\rho|$ of eq.~\ref{e:oneinst}. A similar relation holds true for
multi-instanton configurations. It should be noted that, for a general
field configuration, the action density is positive definite, but the
charge density is not, having contributions from both instantons and
anti-instantons for which the charge has opposite signs. The topological
charge sets a lower bound on the action~\cite{MW:sph}:
\begin{equation}\label{e:lowbound} S\ge\frac{4\pi}{g^{2}}\mid Q\mid
\end{equation} \noindent from which it can be deduced that a single
instanton configuration has action $S_{I}$:
\begin{equation}\label{e:instact} S_{I}=\frac{4\pi}{g^2}\ .
\end{equation} \noindent Clearly an $N$-instanton configuration
generated from eq.~\ref{e:multinst} will have an action $4\pi N/g^2$.

On a space with toroidal boundary conditions, there are {\em no\/} exact
single-instanton solutions to the equations of motion. An argument for
this is given in~\ref{a:torus}. The simplest solution  has two
instantons with their $\rho$ parameters satisfying $\rho_1= -\rho_2$.
For solutions with several  instantons and anti-instantons, the
constraints from the toroidal  boundary conditions will be relatively
less severe. On the lattice the boundary conditions are toroidal, so the
one instanton configuration based on eq.~\ref{e:oneinst} will be only a
metastable field configuration.

Furthermore, the lattice action differs from the continuum action to
order $a$ where $a$ is the lattice spacing. As shown
in~\cite{Sp:overimp}, this implies that instantons are not exact
solutions to the lattice equations of motion. For the conventional
lattice action (as used here) this results in instantons shrinking
slowly under application of the equations of motion.

If a dilute gas of instantons and anti-instantons is a good
approximation, it is possible to deduce the nature of their size
distribution from  renormalisation group arguments---see
~\cite{C:uses,rajaraman,P:strings}.
For the \o3 model this approximation gives 
\begin{equation}\label{e:o3i}
\frac{S}{V}=e^{-4\pi/g^{2}}g^{-4}\int_{0}^{\infty}d\rho\frac{
f(\rho M)}{\rho^3}
\end{equation} 
\noindent where the 4 powers of $1/g$ come from scale and translation
considerations, and the 3 powers of $1/\rho$ in the integrand come from
dimensional analysis. $f$ is the function we wish to determine and $M$ the
cutoff mass. $g^2$ is dimensionless, and as the only dimensionful quantities
are $\rho$ and $M$, then we should replace the bare coupling $g^2$ by the
running coupling $\gb^2(\rho M)$. Renormalisation group theory implies that
for any observable quantity, $\rho M$ and $g$ should enter only in the
combination
\begin{equation}
\frac{1}{g^2}\rightarrow \frac{1}{\gb^2}=\frac{1}{g^2}-\beta_{1} 
\ln (\rho M) + O(g^{2})
\end{equation}
\noindent with $\beta_1$ obtainable from the one-loop $\beta$-function. The
form of $f$ and hence the distribution by size can therefore be deduced. For
\o3, $\beta_{1}=1/2\pi$~\cite{Nov:RG}, (see also~\cite{P:strings}) and gives
a factor of $\rho^2 M^{2}$, implying $f\propto\rho^2 M^2$ to accomodate this,
and giving a distribution that goes as the reciprocal of the instanton size. 
This explains why it is so  difficult to determine the topological charge on
a lattice since the  distribution is peaked at small size where the lattice
modifies the behaviour substantially. However, the large size component of
the  distribution should be amenable to study on a lattice and that is the 
topic we now address.

\section{Cooling and Numerical Lattice Calculations}\label{s:cool}
\begin{figure}[t]
\vspace{3 in}
   \includegraphics{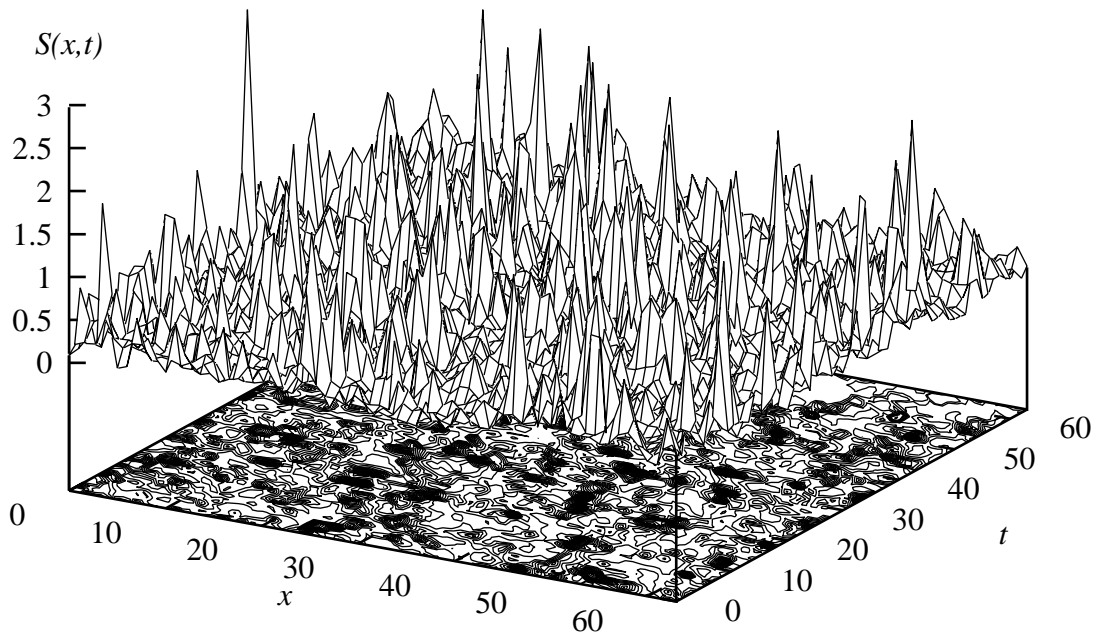}
\vspace{.5 in}
\caption{ 
The action density from an uncooled configuration,
generated at $g^2=0.8$.
\label{f:uncooled}}
\end{figure}

We use lattices of size $64 \times 64$ with 1000 sweeps between 
configurations. Each sweep is composed of 1 heatbath and 9 over-relaxed 
updates of the whole lattice. The action density
distribution on a lattice configuration at $g^2=0.8$ is shown in
fig.~\ref{f:uncooled}. As we found  previously~\cite{US:cool}, it is not
feasible  to derive any useful results from such configurations since 
they have too much short distance fluctuation. They first need to be
cooled.

Cooling, being a local update algorithm, removes the short-range noise
and leaves the longer-range instantons approximately
alone~\cite{Pisa:o3}. Our lattice action has order $a$ effects  that
imply~\cite{Sp:overimp} that an instanton-like solution has an action
close to  the continuum value given in eq.~\ref{e:instact} but which is
reduced  slightly as the size $\rho$ decreases. Thus cooling, which is
equivalent to iteratively implementing the equations of motions, will 
affect the size of an instanton. Protracted cooling on a single,
moderately sized instanton gradually reduces its size to that of one
lattice spacing, and then destroys it. This is familiar from the
plateaux that develop in the action as a configuration is cooled. The
plateaux correspond to instantons being shrunk then destroyed. The
contribution to the action from an instanton is a finite amount which
depends only weakly on the instanton's size, so the action will decrease
by approximately $S_{I}=4\pi/g^2$ each time an instanton is destroyed.
In order to analyse instanton-like features after cooling, we have to 
calibrate the cooling process to take account of such shrinkage under 
cooling.

The cooling we used was an under-relaxed cooling:
\begin{equation}\label{e:urcool}
\phi_{x}'= \alpha\phi_{x} + \phi_{F}(x)
\end{equation}
properly normalised so that $(\phi_x')^2=1$,
with $\phi_{F}(x)$ defined in eq.~\ref{e:phiF}. $\alpha$ is a
parameter which determines the severity of the cooling. The effects of
varying $\alpha$ are discussed in~\cite{US:cool}.

We wish to study the nature of the topological charge distribution, 
in particular the frequency of occurrence of large objects. We 
assume that large scale objects will have a shape similar to that 
of an idealised instanton. This is based on the expectation that 
meta-stable solutions to the classical equations of motion will 
dominate the quantum vacuum, so our strategy is to analyse the 
topography of the cooled vacuum distribution to report on its 
nature.

\begin{figure}[tp]
\vspace{3 in}
   \includegraphics{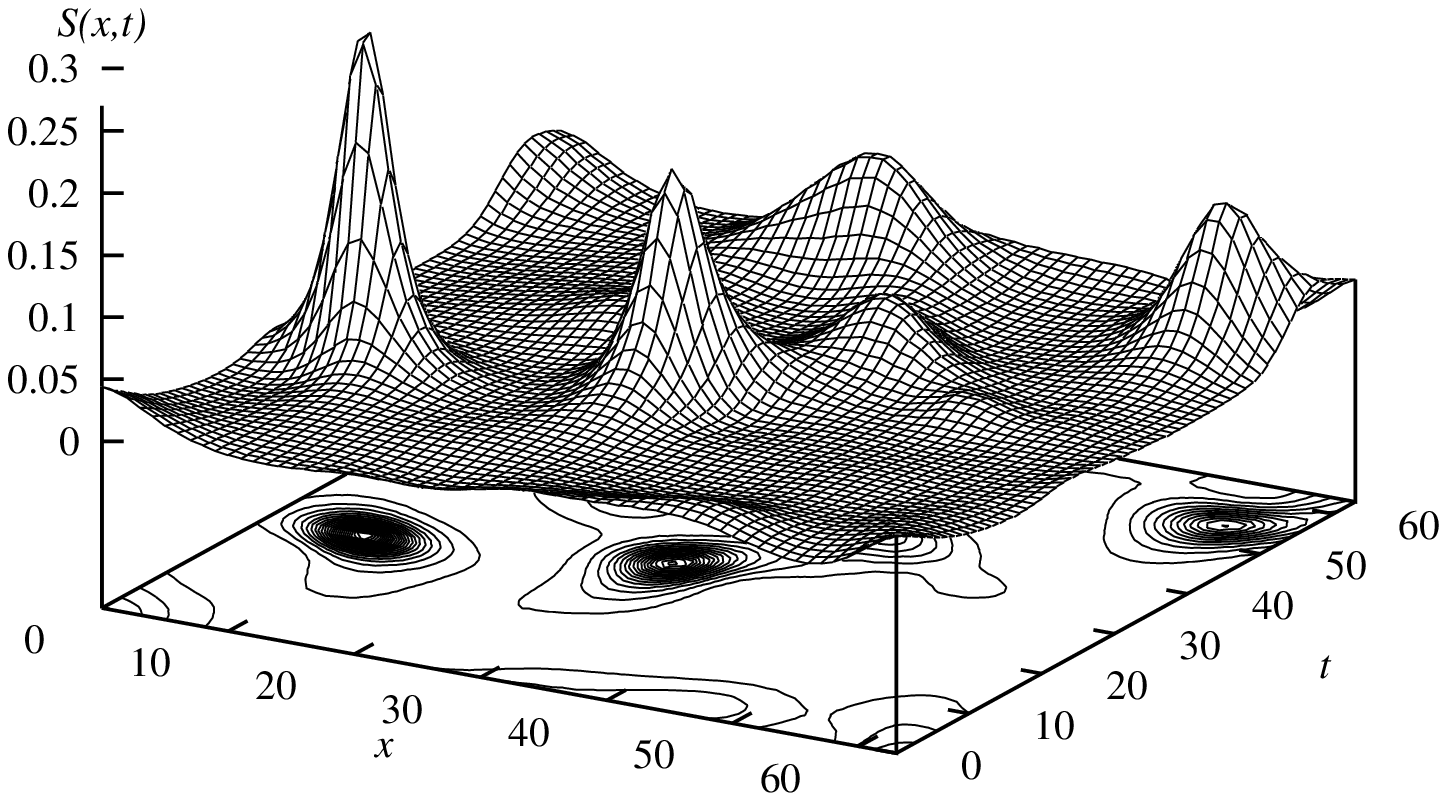}
\vspace{3.5 in}
   \includegraphics{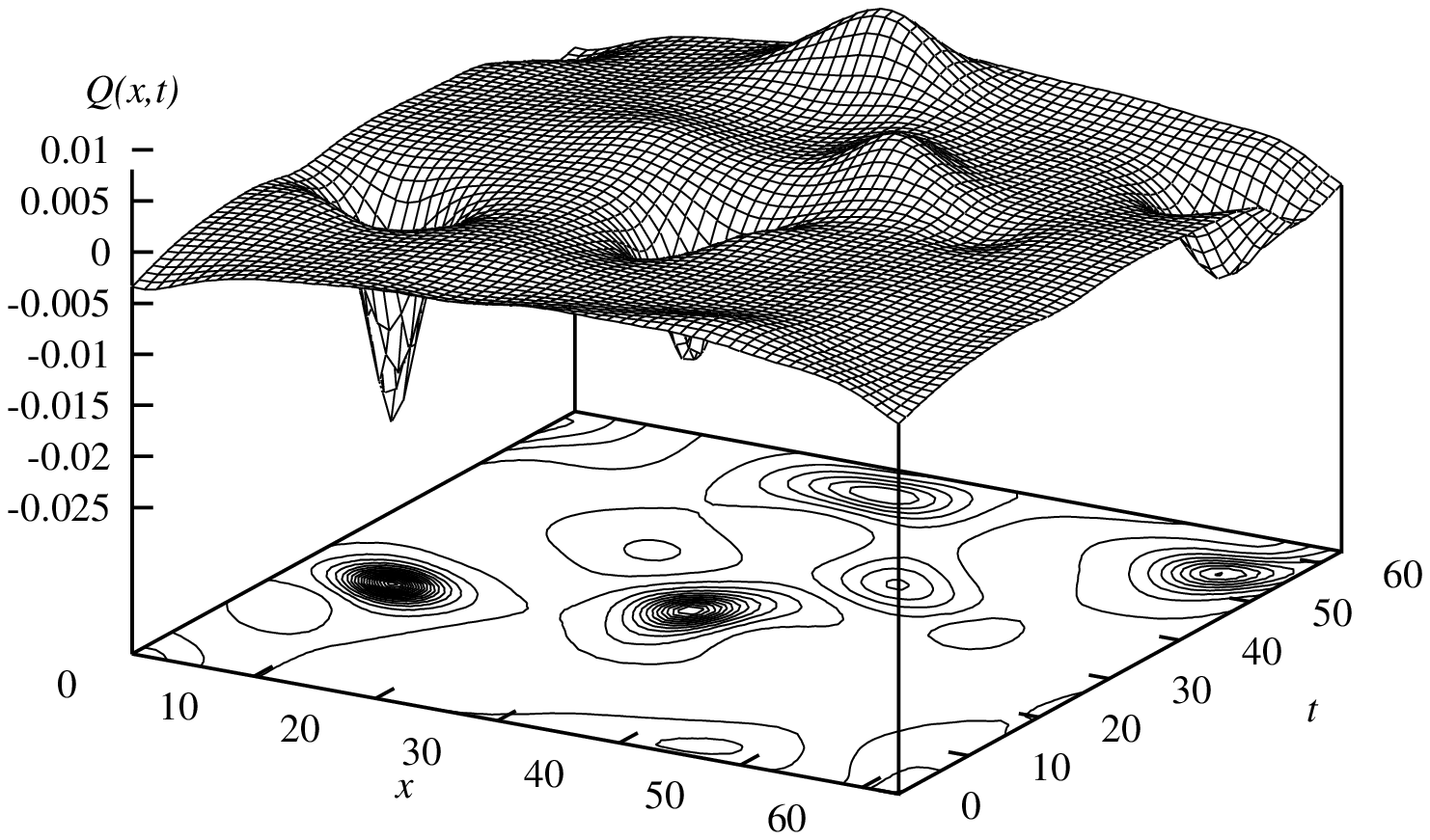}
\vspace{.5 in}
\caption{ 
A cooled configuration at $g^2=0.8$ after 90 cooling sweeps
with $\alpha=2$.\label{f:sandq}}
\end{figure}

If the action after cooling is $S$, then we expect to have of order
$N=S/S_I$ topological objects present. The topological charge
distribution $Q^T_g(x,t)$ can have either sign and so contains more
information in principle than the action density $S(x,t)$.  Since,
however, $|Q^T(x,t)|$ closely tracks $S(x,t)$ after
cooling~\cite{US:cool}---see fig.~\ref{f:sandq}---it is sufficient to
use $S$ alone. This has two advantages: it is easier to calculate
numerically (particularly in more complex theories than \o3) and it
enables us to select regions in a way unbiased by sign so that any
correlation between the distribution of instantons and anti-instantons
can be studied. A further advantage is that from $S(x,t)$ it is
relatively easy to devise a robust algorithm that finds $N$ connected
space-time regions.

For each cooled configuration, we select $N$ connected regions such 
that  each region contains an internal local maximum of the action
density and is composed of all connected sites with an action density
greater than half that maximum. The regions are mutually exclusive and we
select them in order of decreasing  value of their maximum action
density.  The square root of the area of each such region was then taken
to parametrise the actual ``size'' $\rho$ of the region. 

In order to calibrate this algorithm, we first analysed artificial
configurations created with one instanton   generated by
eqs.~\ref{e:phi} and~\ref{e:oneinst}.
A linear relation was found between our calculated size and the input size. 
Returning now to real configurations, we looked at the relation between the
value of the action density at the local maximum and the size assigned to it,
as eq.~\ref{e:sandq} implies, for isolated continuum instantons:
\begin{equation}\label{e:honw}
S(x_{max})=\frac{4}{g^2\rho^2}\ \ .
\end{equation}

\begin{figure}[t]
\vspace{4 in}
   \includegraphics{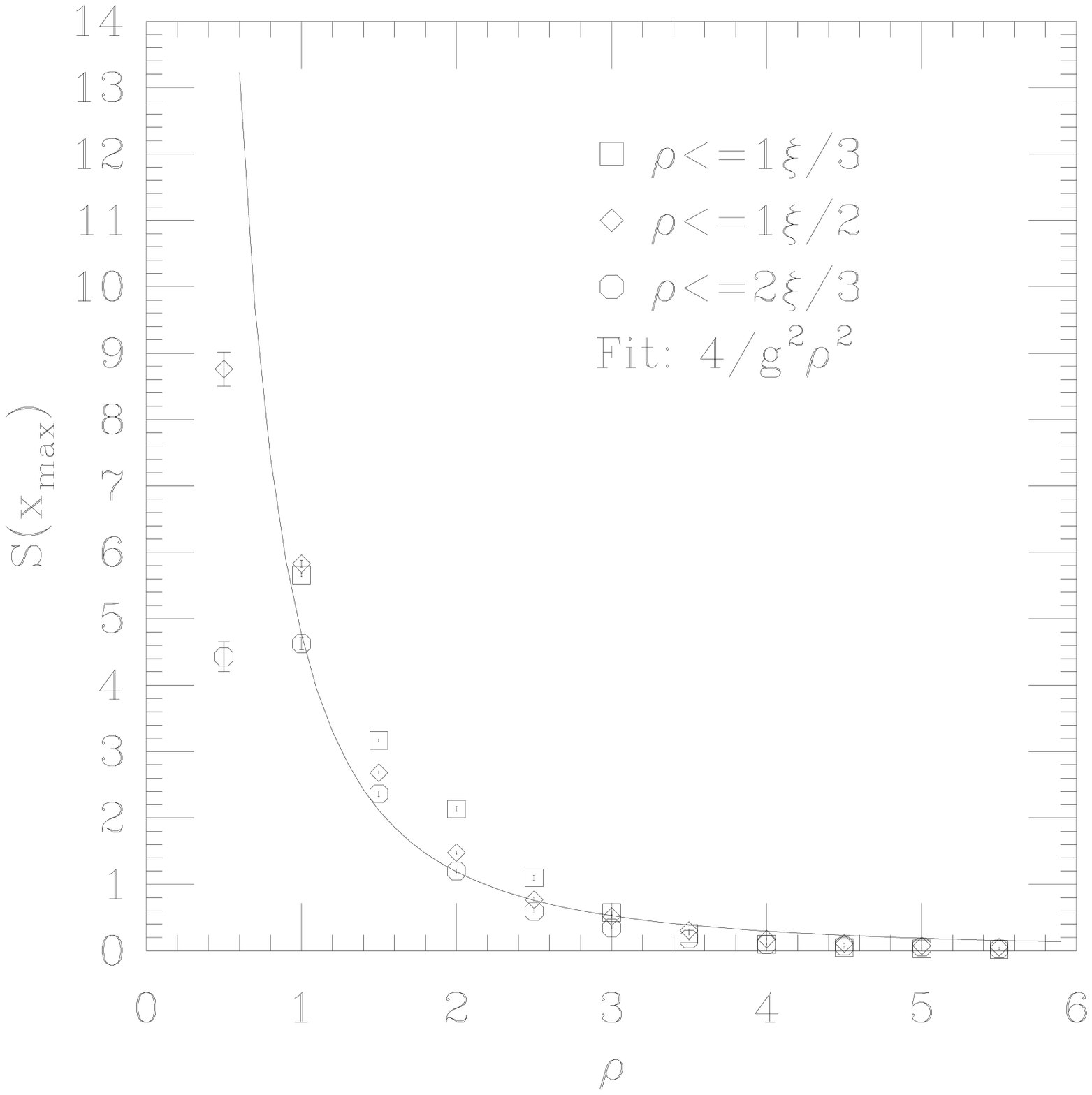}
\vspace{.5 in}
\caption{\label{f:handw}
The relation between the value of the action density at a local
maximum and the instanton `size' we associated with it. For these data
$g^2=0.84$. The curve comes from eq.~\protect{\ref{e:honw}}.}
\end{figure}

We compared the data from our configurations with this predicted
continuum behaviour; the result is shown in fig.~\ref{f:handw}. While
it is clear that our data do {\em not\/} have the exact  inverse-square
behaviour, it is nonetheless encouraging to see that the data lie near
to the curve. Indeed, any dissimilarity between the curve and the data
can be seen as further evidence that the dilute instanton gas is not
the correct model for the vacuum.

Since cooling tends to shrink instantons, it is important to  calibrate
the amount of cooling used and the extent of this effect.  This is
particularly important since we wish to decrease the lattice  spacing
and obtain results for the continuum limit. Our cooling  algorithm is
local (ie. on the scale of 1 lattice spacing) and so will  need to be
adjusted to have the same physical effect as the lattice  spacing is
reduced. We would hope that cooling can be arranged to  have the same
effect in physical distance units.  This may not be possible because 
though the cooling may be adjusted to have similar effects at a given
short distance (in  physical units), it may give different effects at a
larger distance as  the lattice spacing is varied.

In order to mimimise such problems, we  specify the cooling in  terms of
its effect on an instanton.  Thus we cool until an isolated instanton of
a given  physical size is removed (ie shrunk to zero).    Indeed
in~\cite{US:cool}, we determined the amount of cooling to use by
computing the number of steps required, at a given $\alpha$, to remove a
single instanton of a given size generated via eq.~\ref{e:oneinst}. Here
we specify the instanton size in physical units, i.e. in units of the
correlation length $\xi$. We considered the cooling required to remove
such instantons of  size  one third, one half and two thirds the
correlation length, and the results are given in table~\ref{t:mass}.

\begin{table}[th]
\begin{center}
\begin{tabular}{|c|c|c||c|c|c|}\hline
$g^2$ & $m_{eff}$ &$L$&$\rho=\xi/3$&$\rho=\xi/2$&$\rho=2\xi/3$\\ \hline
0.8   & 0.261(1)  & 64   & 28      & 79      & 220 \\
0.84  & 0.318(1)  & 64   & 17      & 45      & 106 \\
1.00  & 0.551(1)  & 64   &  5      & 13      &  24 \\ \hline
\end{tabular}
\caption{\label{t:mass} The measured mass gap and lattice sizes at the
three values of $g^2$ we used, and the number of cooling sweeps required, at
$\alpha=2$ to remove an instanton of size $\rho=\xi/3,\xi/2,2\xi/3$ at
each value of $g^2$.}
\end{center}
\end{table}

Our aim is to derive a distribution of the size of topological
fluctuations which is independent of both the lattice spacing (ie
coupling $g^2$) and the cooling algorithm.  We expect a weak signal
from objects with large physical sizes, because of the effects of the
finite lattice size and because a large instanton has small maximum
action density and so may be swamped by the larger signal from smaller
objects; while at small scales we expect unphysical contributions
from lattice artefacts and from the size cutoff of the lattice to
adversely affect the results. It was our hope that there would be some
intermediate region where sensible results could be obtained.

We analyse 1000 configurations at each value of $g^2$. These correspond 
to a range of correlation lengths of a factor of 2, allowing us to 
study scaling (ie independence of the results on the lattice spacing).
We checked that there was no statistically significant evidence for 
any auto-correlation between these configurations so that they can 
be treated as statistically independent in error analyses. Each 
configuration was cooled  by the amounts shown in table~\ref{t:mass}.
This range of cooling was restricted by the requirement that
the instanton removed for calibration should not be too small 
(thus at $g^2=1.00$  objects of size $\rho\sim\xi/3$ correspond to 
less than 1 lattice spacing). 
At the other cooling extreme, we found that severe cooling to
remove all objects of size $\rho\sim 2\xi/3$ made a rather large 
change to the original configuration and so tended
to modify more  any physics that might be present in the uncooled
configuration. Consequently we took this to be the upper bound on any
useful degree of cooling.

The consistency of our cooling scheme can be judged by measuring the
ratio of the number $N=S/S_I$ of topological objects to the physical
volume $V$. This should be independent of $g^2$ at fixed cooling
level.  Table~\ref{t:vol} shows an approximate independence. This
provides an estimate of the systematic error from a cooling scheme.
We also see a decrease of $N$ with increased cooling at fixed $g^2$,
but this is entirely to be expected.

\begin{table}[th]
\begin{center}
\begin{tabular}{|c|c|c|c|}\hline
$g^2$ & $\rho=\xi/3$ & $\rho=\xi/2$ & $\rho=2\xi/3$ \\ \hline
0.8   & 0.04897(20)  & 0.02745(17)  & 0.01574(14)  \\ 
0.84  & 0.04945(15)  & 0.02749(12)  & 0.01703(11)  \\
1.00  & 0.048889(78) & 0.024663(57) & 0.016261(49) \\\hline
\end{tabular}
\caption{\label{t:vol} The number of objects per unit physical
volume per configuration, $S/VS_I$, at the different values of $g^2$
under differing amounts of cooling.
}
\end{center}
\end{table}

\begin{figure}[t]
\vspace{4 in}
   \includegraphics{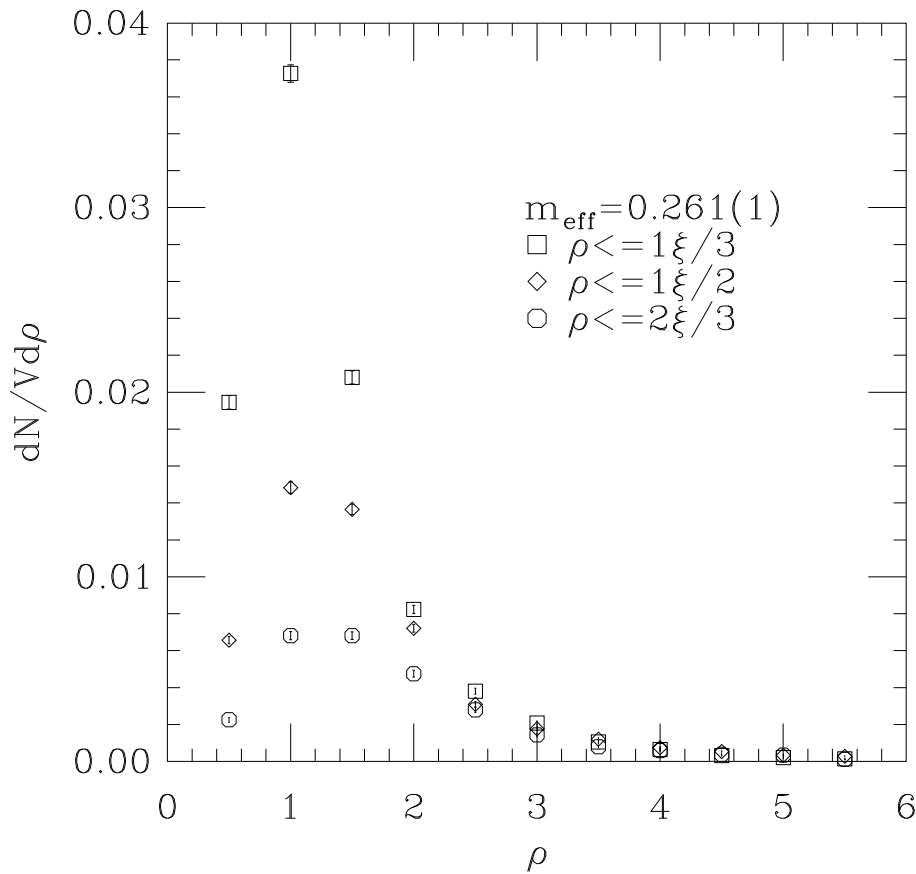}
\vspace{.5 in}
\caption{\label{f:3cool}
Size distribution data obtained from cooling the same 1000
configurations by three different amounts, the details given in
table~\protect{\ref{t:mass}}. For these data $g^2=0.8$. $\rho$ is
given in units of $\xi$.}
\end{figure}

\begin{figure}[t]
\vspace{4 in}
   \includegraphics{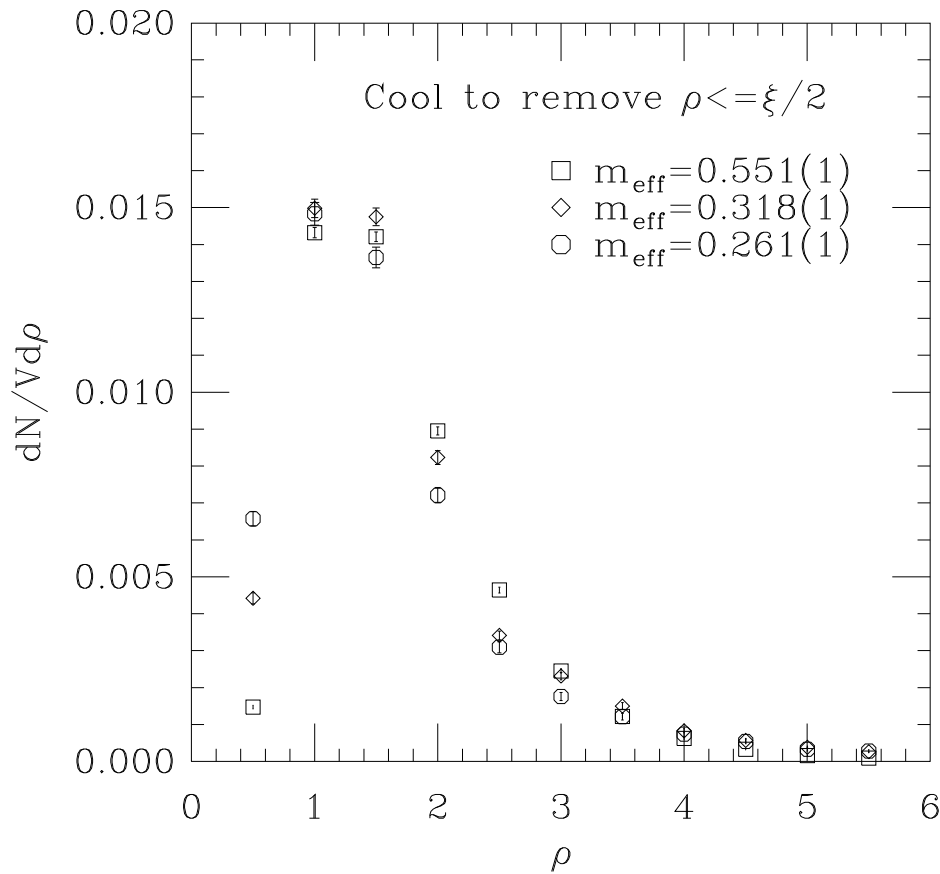}
\vspace{.5 in}
\caption{\label{f:3mass}
Size distribution data obtained from performing the same cooling on
configurations generated at three different couplings, the details given in
table~\protect{\ref{t:mass}}. For these data, the cooling was to
remove objects of size $\rho\lesssim\xi/2$. $\rho$ is given in units of $\xi$.} 
\end{figure}

We now present our results for the size distribution of topological objects 
found by the algorithm introduced above. In figs.~\ref{f:3cool}
and~\ref{f:3mass}, we have plotted $dN/V d\rho$ where both $\rho$ and $V$ are
in units of the correlation length (as are all lengths we quote), and
$N=S/S_I$. 

Let us first discuss the effect of varying the amount of  cooling at a fixed
lattice spacing. We expect that objects of  progressively larger sizes are
modified as the cooling is increased.  One would hope to see that the signal
from large size objects is  independent of cooling. Fig.~\ref{f:3cool} shows
the data obtained from cooling the same 1000 configurations, generated at
$g^2=0.8$, by three different amounts: to remove objects of size
$\rho\sim\xi/3, \xi/2, 2\xi/3$ and smaller. As we expected, we see smaller
objects progressively removed but there is a clear signal that is
cooling-independent in the range $2.5\lesssim\rho\lesssim 4.5$.  For  very
large objects there is virtually no signal.  We performed the same cooling
on configurations generated at other values of $g^2$ and found similar
signals.

We also wish to study the independence of the size distribution  for changing
lattice spacing when the cooling is held the  same.  Fig.~\ref{f:3mass} shows
the data obtained at each of the values of $g^2$ we used, here cooled to
remove objects of size $\rho\sim\xi/2$ and smaller. There is a good agreement
across the range of $g^2$ we used in the range $3\lesssim\rho\lesssim 5.5$;
this coupling-independence is good evidence for a continuum distribution
similar to that shown. For the lesser amount of cooling, we found good
agreement between the results at $g^2=0.8$ and $g^2=0.84$ in a larger range,
but not between these data and those calculated at $g^2=1.00$, where the
cooling corresponded to removing objects of size less than one lattice
spacing. At the greater cooling level, the agreement was significantly less
than for the data shown. We would therefore propose that these are the
outside limits on what useful cooling can be done without adversely affecting
the physics.

\begin{figure}[t]
\vspace{4 in}
   \includegraphics{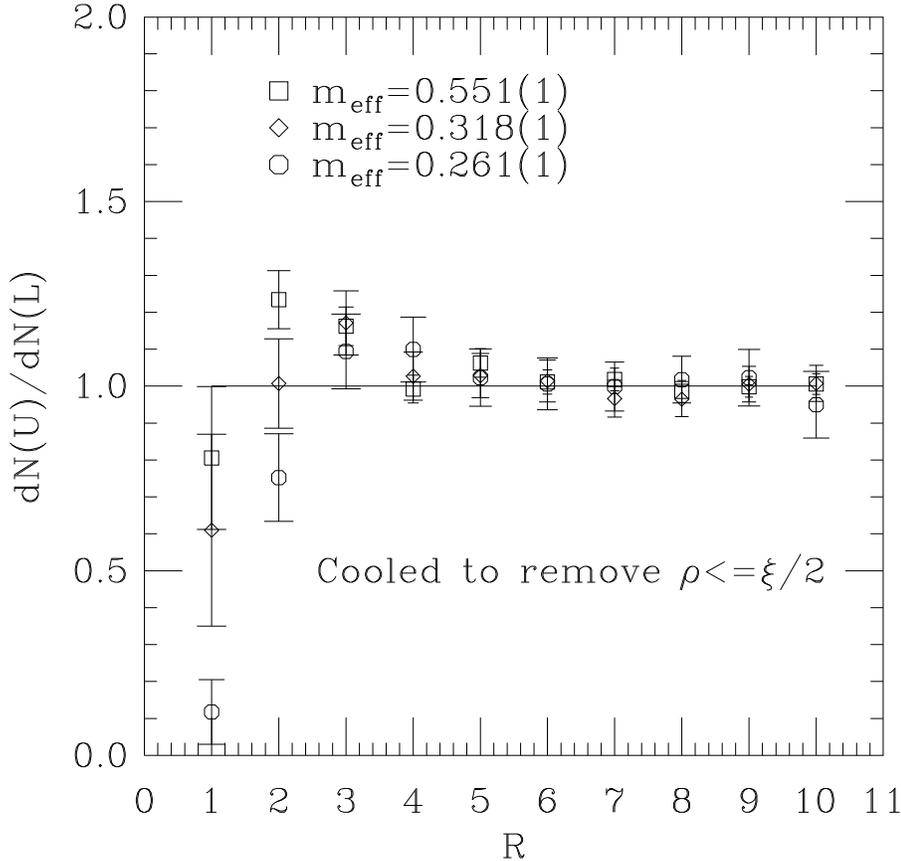}
\vspace{.5 in}
\caption{\label{f:seps}
The ratio of separation distributions between pairs of unlike and of like
objects, calculated on configurations cooled to remove $\rho\lesssim\xi/2$.}
\end{figure}

In an attempt to discover something about instanton--anti-instanton
interactions, we looked at the separations of objects, ie of
instanton--instanton, anti-instanton--anti-instanton and
instanton--anti-instanton pairs. In particular we looked at the ratio
of distance distributions for unlike and like pairs, and at the
average closest separations. The results are presented in
fig.~\ref{f:seps} and table~\ref{t:seps}. We are mainly interested in
small separations, $R$, as we would expect there to be no observable
difference for well-separated objects. As can be seen from
table~\ref{t:seps}, we find that unlike pairs are closer than like
pairs. The decrease in $R$ with correlation length is understandable
as the larger $g^2$ corrersponds to a larger physical volume, and
hence a larger number of objects within that volume.  In
fig.~\ref{f:seps} we plot $dN({\rm U})/dN({\rm L})$, the ratio of
distributions of unlike to like pairs, and find that at small
separations, this ratio is greater than one, indicating that there
are indeed more unlike pairs at smaller distances. At {\em very\/}
small distances any possible signal is destroyed by the cooling
process, and at large separations there would seem to be as many like
as unlike pairs, as intuition would suggest. We found that the number
of like and unlike pairs were consistent across all coolings and the
three values of $g^2$ we used. This is clearly an area of some
interest in which more work than our preliminary study could be
undertaken.

\begin{table}[h]
\begin{center}
\begin{tabular}{|c|c|c|c|c|}\hline
$g^2$ & Separation & $R$ & \# discounted & \# used \\ \hline\hline
$0.8$       & I-I & 4.371(62) & 73 & 927  \\
            & A-A & 4.426(64) & 72 & 928  \\
            & I-A & 3.331(33) & 21 & 979  \\
            & U/L & 0.757(18) &    &      \\ \hline
$0.84$      & I-I & 3.914(54) & 7  & 993  \\
            & A-A & 3.828(57) & 8  & 992  \\
            & I-A & 2.781(24) & 0  & 1000 \\
            & U/L & 0.718(16) &    &      \\ \hline
$1.00$      & I-I & 2.681(26) & 0  & 1000 \\
            & A-A & 2.612(27) & 0  & 1000 \\
            & I-A & 2.090(16) & 0  & 1000 \\
            & U/L & 0.789(14) &    &      \\ \hline
\end{tabular}
\caption{\label{t:seps} The closest separation of objects at the different values of $g^2$
under differing amounts of cooling. The entries denoted `U/L' are the ratios
of the I-A separations to the average of the I-I and A-A. The 4th and 5th
columns are, respectively, the number of configurations for which a
calculation was not possible (ie those containing only one or fewer
(anti-)instantons) and the number of configurations used in the actual
measurement.}
\end{center}
\end{table}

We compare our method of analysis of the large scale topological 
objects with that used by other workers.
In~\cite{Neg:hadrons} the size of \su2 instantons was estimated
by calculating the topological charge density correlation function
\begin{equation}
f(x)=\sum_yQ(y)Q(x+y)
\end{equation}
\noindent with the sum over the whole lattice. Then $f(x)$ is fitted for
each instanton at a particular value of $\rho$
using a convolution of an analytic continuum expression similar to
eq.~\ref{e:sandq}:
\begin{equation}
Q_{\rho}(x)=\frac{6}{\pi^2\rho^4}\left(\frac{\rho^2}{x^2+\rho^2}\right)^4
\end{equation}

Elsewhere, in~\cite{PV:su2size}, \su2 instanton size data are obtained from
two methods: first a simple extraction of a size from the continuum
single-instanton relation
\begin{equation}
S_{max}=\frac{48}{g^2}\frac{1}{\rho^4}
\end{equation}
\noindent and a second method which measured the extent of the
instanton only in the lattice axis directions, and then took the
average of these numbers as the size.

Different methods can give different results, particurarly for  \o3,
since the vacuum is not well approximated by a dilute gas of 
instantons. We believe that our method of assigning ``instantons'' is more
 robust and works well even if the instantons are not
well-separated. Our method works for  moderately cooled configurations
and does not rely upon the configurations having been cooled to such an
extent that they are no longer representative of the pre-cooled physics.
Furthermore, the methods used in~\cite{PV:su2size} rely explicitly on
the instantons having a circular geometry, just as those generated by
eq.~\ref{e:oneinst} do; it is clear from fig.~\ref{f:sandq}, and from
the results shown in~\cite{US:cool}, that, in \o3 at least, artificially
generated isolated instantons are not representative of  those obtained
from a simulation of the vacuum.

\section{Conclusions} 

In this paper we have extended the work begun in~\cite{US:cool}, and
shown that it is possible to obtain a distribution of large size
topological objects that is, within some size range, independent of
the cooling performed. We have put forward the criterion for an
optimum cooling to be such that objects of size ${\cal O}(2a)$, ie
twice the lattice spacing be removed by the cooling process. At our
values of $g^2$, this corresponds to objects with size of the order of
half the correlation length be removed. Furthermore, we find results
that can be interpreted as applicable to the continuum distribution,
as they are independent over a range of lattice spacing varying by a
factor of two.
 
We confirm, by non-perturbative study, that the distribution of 
size of topological objects is peaked at small size. Roughly our 
results are that
\begin{equation}\label{e:dndr}
\frac{1}{V}\,\frac{dN}{d\rho}\sim\frac{1}{\rho^3}
\end{equation}
\noindent (with $N=S/S_I$). This contrasts with the inverse size 
distribution expected for a dilute gas of instantons.

We have also shown, in figs.~\ref{f:sandq} and~\ref{f:handw}, that the
dilute instanton gas is not a good model for the \o3 vacuum at the
values of $g^2$ we used, and made a preliminary study of the relative
distributions of instanton and anti-instanton separations, and found
them to be correlated to the extent that unlike pairs exist at smaller
separations that like pairs of objects.

\appendix
\section{Instantons on a torus}\label{a:torus}
In this appendix we present an argument that it is not possible to
place a single instanton on a torus.

If we write the action and topological charge in terms of
$\omega,\overline{\omega}$ we get~\cite{RR:torus}: 
\begin{equation}\label{e:somega}
S(\omega)=\frac{4}{g^2}\int\frac{d^2x}{(1+\omega\overline{\omega})}
\left(\partial_z\omega\partial_{\overline{z}}\overline{\omega}+
\partial_z\overline{\omega}\partial_{\overline{z}}\omega\right)
\end{equation}
\noindent and
\begin{equation}\label{e:qomega}
Q(\omega)=\frac{1}{\pi}\int\frac{d^2x}{(1+\omega\overline{\omega})}
\left(\partial_z\omega\partial_{\overline{z}}\overline{\omega}-
\partial_z\overline{\omega}\partial_{\overline{z}}\omega\right)
\end{equation}
\noindent which lead us to the following expression for $S$:
\begin{equation}\label{e:sqomega}
S(\omega)=\frac{4\pi}{g^2}|Q(\omega)|+\frac{8}{g^2}\int\frac{d^2x}
{(1+\omega\overline{\omega})}\mid\partial_{\overline{z}}\omega\mid^2
\end{equation}

Instantons are locally stable, finite-action solutions to the
Euclidean equations of motion, and as such are given by
\begin{equation}\label{e:eqmotion}
\partial_{\overline{z}}\omega=0
\end{equation}
\noindent and we therefore require $\omega$ to be an analytic
function.

If we now take a rectangular region, $R$, of ${\Bbb R}^2$ and impose toroidal
boundary conditions, then ($\partial R$ denoting the boundary of $R$)
\begin{equation}\label{e:contourint}
\oint_{\partial R}\omega(z)dz=0
\end{equation}
\noindent as there is a pairwise cancellation of the line integral
along opposite sides of $R$. Cauchy's theorem then implies that 
\begin{equation}\label{e:cauchy}
\sum{\rm residues}=0
\end{equation}
\noindent and thus that one cannot have a single-instanton solution to
the equations of motion on a torus (as its residue would be
identically zero) but multi-instanton solutions are possible. As shown
in~\cite{RR:torus}, a periodic solution of eq.~\ref{e:eqmotion} on $R$
can be expressed in terms of the Weierstrass $\sigma$ function.

\end{document}